\documentclass[runningheads]{llncs}
\usepackage[T1]{fontenc}
\usepackage{graphicx}
\usepackage{xcolor}
\usepackage{enumitem}
\usepackage{hyperref}
\begin{document}
\title{LLMs+Graphs: Toward Graph‑Native, Synergistic AI Systems}
\author{Arijit Khan\inst{1}\orcidID{0000-0002-7312-6312} \and
Longxu Sun\inst{2}\orcidID{0000-0002-7465-9153} \and
Xin Huang\inst{2}\orcidID{0000-0002-3650-0301}}
%
%\authorrunning{A. Khan et al.}
\institute{Bowling Green State University, Ohio, USA \\
\email{arijitk@bgsu.edu}\\
Hong Kong Baptist University, Hong Kong, China\\
\email{\{cslxsun,xinhuang\}@comp.hkbu.edu.hk}}
\maketitle              % typeset the header of the contribution
\begin{abstract}
Large Language Models (LLMs) have advanced rapidly, but their limitations in structured and multi‑hop reasoning underscore the need for graph‑native, synergistic artificial intelligence (AI) systems. Graph‑structured data underpins critical applications across social, biological, financial, transportation, web, and knowledge domains, making it essential to understand how LLMs can leverage graph computation for grounded, context‑rich inference. Three complementary synergies are emerging: LLMs augmented with graph computation for retrieval and reasoning; bidirectional integration between LLMs and knowledge graphs (KGs), where LLMs support KG construction and curation while KGs enforce semantic constraints and factual consistency; and AI agents strengthened by graph algorithms for planning, decision making, and multi‑step reasoning. In parallel, LLMs introduce new capabilities for graph data management and graph machine learning (ML) through natural language interfaces and hybrid LLM-graph neural network (GNN) pipelines. This tutorial synthesizes the algorithms, systems, and design principles driving these converging directions, offering data science and data mining researchers a unified perspective on integrating LLMs, graph data management, graph mining, graph ML, and agentic computation into next‑generation graph‑native AI systems.
%
%\keywords{Large Language Models \and Graph Foundation Models \and Retrieval-augmented Generation \and Knowledge Graphs \and AI Agents.}
\end{abstract}
\vspace{-2mm}
\section{Goals and Objectives}
LLMs have quickly become a primary interface for data‑intensive applications, reshaping how users query, explore, and manage information \cite{FernandezEFKT23,HalevyCFFNW23,Tan23,zhou2025surveyllmtimesdata}. Their emergence coincides with a growing need to reason over graph‑structured data, e.g., knowledge graphs, heterogeneous networks, and text‑attributed graphs that encode rich semantics and relational structure. At the same time, existing graph neural networks (GNNs) and other deep graph models rely heavily on task‑specific supervision and struggle to generalize across the diverse and rapidly evolving landscape of graph applications \cite{ShangH25,abs-2505-15116}. This has motivated the development of graph foundation models, which aim to unify LLMs’ semantic and reasoning capabilities with graph‑native computation to support broad, transferable graph intelligence \cite{LiuYLCLZBFSYS25,WangZLC025}. As modern applications increasingly depend on complex graph data, understanding how LLMs can construct, curate, and leverage these structures--and how graphs can in turn ground and enhance LLM reasoning--has become a central challenge for the data mining community.

Concurrent advances in LLM-based AI agents have heightened the importance of graph-centric architectures. Agents rely on planning graphs, memory graphs, tool-calling graphs, and KG-grounded reasoning to achieve reliable planning, multi-step decision making, and long-term coherence \cite{abs-2506-18019}. At the same time, agents are beginning to reshape core data management tasks--including KG construction, graph reasoning, query formulation and optimization--creating a mutually reinforcing frontier between agentic computation and graph data systems \cite{LLFT26,gds,0004BGCYMPFMIMP25}. This tutorial is motivated by the convergence of these trends: adapting LLMs' semantic and generalization capabilities to graph tasks, advancing graph foundation models, and leveraging graph structures to enable robust, interpretable, and scalable agentic behavior. For the PAKDD community, this convergence presents a timely opportunity to shape the next generation of graph-native, synergistic AI systems.

\textbf{Previous Offering and Related Tutorials.}
We have not presented this tutorial at any prior venue, and it differs substantially from existing offerings. While prior tutorials focus narrowly on LLM–GNN alignment for graph learning \cite{0001RTYC24}, LLM+RAG \cite{llmrag}, KG reasoning \cite{liu2024new}, or LLM+KG integration \cite{Ma000W25}, our tutorial is the first to provide a unified, end‑to‑end treatment of the full synergy between LLMs and graph data. It covers both directions--LLMs for graphs and graphs for LLMs--while simultaneously integrating knowledge graphs, graph‑structured data, and LLM‑based agents into a single coherent framework. This breadth and unification constitute the core novelty of our proposal.
\vspace{-2mm}
\section{Target Audience}
This tutorial is intended for participants working in the broader area of large language models, graph foundation models, graph learning, graph data mining, knowledge-augmented models, agentic AI, and graph AI systems from both academia and industry.
Familiarity with basic LLM techniques would be helpful. %The tutorial is designed for 40\% novice, 30\% intermediate, and 30\% expert  participants, to maintain a balance across overview, applications, and technical contents.
\vspace{-2mm}
\section{Outline} 
%The lecture-style tutorial will provide essential background on LLMs, foundation models, retrieval-augmented generation (RAG), knowledge graphs, and AI agents, along with a selective survey of recent advances in LLMs+Graphs and emerging graph‑native AI systems. The intended length of our tutorial is 3 hours.
\begin{small}
\begin{verbatim}
1 Introduction (30 minutes)
  - Large Language Models, Foundation Models, AI Agents,
    Retrieval-augmented Generation, Knowledge Graphs
2 LLMs for Graphs (20 minutes)
3 Graphs for LLMs (20 minutes)
4 Knowledge Graphs for LLMs (20 minutes)
5 LLMs for Knowledge Graphs (20 minutes)
6 Graphs for AI Agents (20 minutes)
7 AI Agents for Graphs (20 minutes)
8 Future Directions (30 minutes)
\end{verbatim}
\end{small}
%
%\vspace{-5mm}
\section{Description of Topics}
%
%\vspace{-2mm}
%\subsection{Background}
%We begin with core background concepts: Large Language Models provide the foundation for open‑ended language understanding and generation. Foundation Models broaden this capability across multiple modalities--such as text, images, audio, and video--to create general‑purpose backbones for diverse tasks \cite{ZhouLLYLWZJYHPLWLXXPYS25}. AI agents build on these models by coordinating planning, tool use, and multi-step decision making to accomplish complex goals \cite{abs-2503-21460}. However, their reasoning often lacks the precise, domain‑specific knowledge required for reliability. Knowledge Graphs address this gap by supplying structured, semantically grounded representations of entities and relations \cite{WeikumDRS21}. Retrieval‑augmented Generation--including variants such as Graph RAG and KG-RAG--then integrates this curated knowledge directly into the inference process, producing more accurate, context‑aware, and trustworthy outputs for domain‑intensive applications \cite{KLZZZ25}.
%
\vspace{-2mm}
\subsection{LLMs for Graphs}
LLMs are increasingly being used in many graph data management, mining, and ML problems \cite{0001LW0S0Y24,ShangH25,JinLHJJH24,RenTYCH24}. \textbf{
$\bullet$ Graph Querying.} Acting as \emph{predictors}, the language understanding capacity of LLMs makes them suitable for processing natural language questions (NLQs) over structured graphs~\cite{ganesan2024llm,neo4j,Liu0GWXJ24}; they enable natural language interfaces for graph querying, where they translate NLQs into executable GraphQL or Cypher queries~\cite{ganesan2024llm,neo4j,Liu0GWXJ24} and power systems such as Neo4j’s NLQ2Cypher pipeline~\cite{neo4j}. \textbf{$\bullet$ Graph Mining.} In graph mining, LLMs again serve as \emph{predictors}, using their reasoning and code generation abilities to extract graph properties and perform tasks such as graph classification, shortest path computation, cycle detection, and subgraph matching~\cite{liu2023one,zhang2024graphtranslator,huang2024can}, as demonstrated by GraphWiz~\cite{chen2024graphwiz}. \textbf{$\bullet$ Graph Learning.} LLMs enhance graph learning by serving as \textit{enhancers} and supporting \textit{GNN-LLM alignment}. They enable zero-shot reasoning over text-attributed graphs and motivate graph foundation models~\cite{liu2023one,zhang2024graphtranslator}. Unified architectures such as Liu et\,al.'s model~\cite{liu2023one} handle node classification, link prediction, and related tasks without task-specific designs. 
\vspace{-2mm}
\subsection{Graphs for LLMs}
Graph‑based retrieval‑augmented generation (Graph RAG) improves LLM accuracy by supplying structured, relationship‑rich context instead of relying only on flat text retrieval. Microsoft's GraphRAG builds a document‑derived graph, organizes it into communities, and produces community-level summaries that offer coherent context for downstream reasoning~\cite{edge2024local}. ArchRAG advances this paradigm by enriching user queries with attributed subgraph communities from an external corpus and introducing an index that improves retrieval relevance and efficiency~\cite{abs-2502-09891}. Zhou et al. provide an extensive empirical comparison of recent graph‑based RAG methods, showing how graph structure enables more faithful and contextually grounded LLM outputs~\cite{ZhouSSWWHZLLMF25}.
\vspace{-2mm}
\subsection{Knowledge Graphs for LLMs}
We categorize the role of KGs in enhancing LLMs as follows \cite{pan2023unifying,Ma000W25}. \textbf{$\bullet$ Background Knowledge.} KGs provide structured facts that improve LLM reasoning by aligning subgraphs with text for joint training~\cite{yasunaga2022deep}. InfuserKI selectively integrates KG facts to reduce forgetting~\cite{wang2024infuserki}, while KG-Adapter and GAIL inject KG structure through efficient fine-tuning~\cite{tian2024kg,zhang2024gail}. GRAG retrieves the top-$k$ relevant subgraphs and aligns graph and text embeddings~\cite{hu2024grag}, and KG-RAG retrieves curated KG triples for fact-grounded reasoning across QA, recommendations, and data management~\cite{ma2025llmkg4qa,0002F0LMWY25,MaKGRAG4SM25,ma2026costefficientragentitymatching}. \textbf{$\bullet$ Reasoning Guidelines.} KGs guide LLM reasoning by supplying candidate subgraphs or shaping each reasoning step. EtD extracts fine-grained KG facts for knowledge-enhanced prompts~\cite{liu2024explore}, and GCR encodes KGs as tries for graph-constrained decoding~\cite{luo2024graph}. Online methods such as LLM-ARK and ToG support sequential KG-guided decisions~\cite{huang2023llm,sun2023think}, while agent-based systems like KG-Agent and ODA integrate KG tools and memory for iterative reasoning~\cite{JiangZZS0ZW25,sun2024oda}. \textbf{$\bullet$ Refiners \& Validators.} KGs refine and validate LLM outputs by filtering incorrect answers and grounding responses in factual structure. ACT‑Selection and KG‑Rank re‑rank candidates using KG types and medical KGs~\cite{salnikov2023answer,yang2024kg}, and KGR verifies factual statements in generated text~\cite{guan2024mitigating}. EFSUM summarizes KG evidence for zero‑shot QA~\cite{ko2024evidence}, InteractiveKBQA enables iterative KG‑guided correction~\cite{xiong2024interactive}, and LPKG fine‑tunes LLMs with KG‑based planning data to improve complex reasoning~\cite{wang2024learning}.

\vspace{-2mm}
\subsection{LLMs for Knowledge Graphs}
LLMs augment KGs through knowledge extraction, completion, embedding, querying, analytics, and domain applications. \textbf{$\bullet$ KG Creation.} Multi-modal LLMs extract entities, relations, and facts from heterogeneous sources such as text, images, and tables~\cite{DengSLWY22,abs-2305-15321}. They also support discovery, typing, linking, and end-to-end KG construction~\cite{WaddenWLH19,JoshiLZW19}. \textbf{$\bullet$ KG Completion.} LLMs improve link prediction by combining textual signals with KG facts~\cite{abs-1909-03193}. Recent models directly generate missing entities in KG triples~\cite{SaxenaKG22}. \textbf{$\bullet$ KG Embedding.} LLMs enrich KG embeddings by integrating textual semantics, as in KEPLER and K-BERT~\cite{WangGZZLLT21,LiuZ0WJD020}. Multi-modal encoders further extend KG embeddings with image and graph information~\cite{abs-2206-13163}. \textbf{$\bullet$ KG Querying.} LLMs interpret natural language questions, extract entities and relations, and support KG-grounded reasoning~\cite{YasunagaRBLL21}. They also translate NLQs to SPARQL~\cite{AvilaVFC24} and integrate KG facts into retrieval-augmented QA~\cite{abs-2306-04136,abs-2309-11206}. \textbf{$\bullet$ KG Analytics.} LLMs assist with graph reasoning tasks such as computing sizes, degrees, and connectivity, supported by prompting-based methods for natural language graph problems~\cite{abs-2308-07134}. \textbf{$\bullet$ Domain-specific KG Applications.} LLM-KG synergy benefits healthcare, biomedical science~\cite{SungLYJKK21}, education~\cite{LHDNW24}, e-commerce~\cite{RCHd23}, and spatio-temporal analysis~\cite{abs-2310-10196}.

\vspace{-2mm}
\subsection{Graphs for AI Agents}
Recent work shows that key elements of agent performance are often represented more effectively with graphs, motivating agents that interact with diverse graph types. \textbf{$\bullet$ Task Planning Graphs.} Graphs support task reasoning, decomposition, and decision search. Knowledge graphs and Graph-of-Thought provide multi-hop context~\cite{WWS24,BestaBKGPGGLNNH24}, task-dependency graphs structure sub-task relations~\cite{WuSSSWZFCCXL24}, and state-space graphs guide sequential decisions~\cite{LeurentM20}. \textbf{$\bullet$ Task Execution Graphs.} Execution improves when agents use graphs to organize tools and environments. Tool-calling graphs model function dependencies for efficient sequencing~\cite{abs-2403-00839}, and environment-interaction graphs capture relationships among agents and entities~\cite{GalliciMM23}. \textbf{$\bullet$ Memory Graphs.} Memory benefits from graph structures that expose relationships for retrieval and long-term reasoning. Recent work uses hierarchical knowledge graphs for structured memory~\cite{abs-2412-05547}, graph-based RAG for accurate retrieval~\cite{abs-2502-14902}, and dynamic graphs that evolve with new experiences~\cite{GutierrezS0Y024}. \textbf{$\bullet$ {Multi-agent Interaction Graphs.}} Multi-agent systems rely on structured communication, and coordination graphs provide clear pathways for agent interaction~\cite{abs-2503-07675}.

\vspace{-2mm}
\subsection{AI Agents for Graphs}
Agents can address many graph data management and mining tasks.  \textbf{$\bullet$ Graph Reasoning.} Agents outperform stand‑alone LLMs by decomposing problems into structured steps and invoking graph tools, enabling reliable execution of tasks such as shortest paths, cycles, triangle counting, maximum flow, PageRank, centrality, community detection, and node similarity~\cite{abs-2410-05130,gds,yuan-etal-2025-ma}. \textbf{$\bullet$ Text2Cypher \& Text2SPARQL.} Agentic systems improve semantic parsing by retrieving schema elements, validating intermediate queries, and iteratively repairing errors~\cite{abs-2511-08274,abs-2507-16971}. \textbf{$\bullet$ KG Construction.} Agents enhance KG construction by using domain‑aware instructions and iterative tool‑based refinement to extract relations and handle complex reasoning more accurately than stand‑alone LLMs~\cite{NingL24}.

\vspace{-2mm}
\subsection{Future Directions}
We conclude by summarizing open challenges and opportunities for data mining. 
\textbf{$\bullet$ Unification of LLM{+}KG{+}Vector DB \& NeuroSymbolic AI.} Future systems must tightly integrate symbolic KGs, neural LLMs, vector retrieval, and rule-based reasoning into unified architectures that support consistent cross-modal inference and robust neuro-symbolic decision making. 
\textbf{$\bullet$ KG-based Agentic Memory.} Agents require scalable KG-structured memory that can store, update, and retrieve long-term knowledge to support reliable multi-step reasoning. 
\textbf{$\bullet$ Unifying Long Context with RAG.} Combining long-context models with KG-guided RAG demands mechanisms that balance extended attention with structured retrieval to prevent drift and hallucination. 
\textbf{$\bullet$ Explainability.} LLM+KG+agent systems need principled methods that expose graph-grounded evidence and trace decision pathways for transparent and trustworthy reasoning. 
\textbf{$\bullet$ Security and Privacy.} Protecting KG content, agent memory, and retrieval pipelines requires defenses against leakage, poisoning, unauthorized inference, and adversarial manipulation.

\vspace{-2mm}
\section{Biography}

\noindent\textbf{Arijit Khan} %(\url{https://www.cs.bgsu.edu/arijitk/index.html}) 
is an Associate Professor at Bowling Green State University (Ohio, USA). 
%His PhD is from University of California, Santa Barbara, USA, and he did a post-doc in the Systems group at ETH Zurich, Switzerland. He has been an assistant professor at Nanyang Technological University, Singapore and an associate professor at Aalborg University, Denmark. 
His research is on data management and AI for the emerging problems in large graphs. He is an IEEE senior member and an ACM distinguished speaker. %Arijit is the recipient of the IBM Ph.D. Fellowship (2012-13), a VLDB Distinguished Reviewer award (2022), two SIGMOD Distinguished PC awards (2024, 2025), and a KDD outstanding PC award (2025). He is the author of a book on uncertain graphs and over 110 publications in top venues including SIGMOD, PVLDB, TKDE, ICDE, ICLR, KDD, EMNLP, SDM, USENIX ATC, EDBT, Web Conference (WWW), WSDM, CIKM, TKDD, VLDB Journal, and SIGMOD Record. Dr Khan served/is serving as an associate editor of TKDE and TKDD, co-Editor-in-Chief of Knowledge Engineering Review (KER), CIKM short paper track co-chair 2024, ICDE demonstration paper track program co-chair 2025, KDD 2025 PhD consortium track program co-chair, and Australasian Database Conference (ADC) 2025 PC co-chair. 
More info at \url{https://www.cs.bgsu.edu/arijitk/index.html}.

\noindent\textbf{Longxu Sun} %(\url{https://www.comp.hkbu.edu.hk/~cslxsun/}) 
is a postdoctoral research fellow in the Department of Computer Science, Hong Kong Baptist University (HKBU). %She has multiple publications in VLDB, ICDE, and TKDE. 
Her research interests include graph mining, community search, and knowledge graph-based community search for Graph RAG. %She earned her PhD from Hong Kong Baptist University in 2024.
More info at \url{https://www.comp.hkbu.edu.hk/~cslxsun/}.

\noindent\textbf{{Xin Huang}} is an Associate Professor at the Department of Computer Science, Hong Kong Baptist University. %He received the Ph.D. degree from the Department of Systems Engineering and Engineering Management at the Chinese University of Hong Kong in 2014. 
His research interests include graph data management, graph mining and visualization, social network analysis, and privacy-aware computing. %He has published in refereed conferences and journals including SIGMOD, PVLDB, ICDE, WWW, KDD, NeurIPS, AAAI, IJCAI, TKDE, TKDD, TMC, TNNLS, and VLDBJ. He is an associate editor of Data Science and Engineering (DSE) and World Wide Web Journal (WWWJ).
More info at \url{https://www.comp.hkbu.edu.hk/~xinhuang/}.
\bibliographystyle{splncs04}
\bibliography{sample-base}

\end{document}